\documentclass{aa} % for a referee version
\usepackage{natbib}
\usepackage{graphicx}
\usepackage{color}
\usepackage{txfonts}
\begin{document} 

   \title{Near- to mid-infrared spectroscopy of the heavily obscured AGN LEDA~1712304 with AKARI/IRC}

   %\subtitle{I. Overviewing the $\kappa$-mechanism}

  \author{T. Tsuchikawa, 
          \inst{1}
          H. Kaneda,\inst{1} S. Oyabu,\inst{1} T. Kokusho,\inst{1} K. Morihana,\inst{1} H. Kobayashi,\inst{1}\\ M. Yamagishi\inst{2} \and Y. Toba\inst{3,4,5} 
          }

   \institute{Graduate School of Science, Nagoya University, 
              Furo-cho, Chikusa-ku, Nagoya, Aichi 464-8602, Japan\\
              \email{tsuchikawa@u.phys.nagoya-u.ac.jp}
         \and
             Institute of Space and Astronautical Science, Japan Aerospace Exploration Agency, 3-1-1 Yoshinodai, Chuo-ku, Sagamihara, Kanagawa, 252-5210, Japan
	\and
		Department of Astronomy, Kyoto University, Kitashirakawa-Oiwake-cho, Sakyo-ku, Kyoto 606-8502, Japan
	\and
		Academia Sinica Institute of Astronomy and Astrophysics, 11F of Astronomy-Mathematics Building, AS/NTU, No.1, Section 4, Roosevelt Road, Taipei 10617, Taiwan
	\and
		Research Center for Space and Cosmic Evolution, Ehime University, 2-5 Bunkyo-cho, Matsuyama, Ehime 790-8577, Japan
             }

   \date{Received ** **, 2019; accepted ** **, 2019}

\abstract{Although heavily obscured active galactic nuclei (AGNs) have been found by many observational studies, the properties of the surrounding dust are poorly understood. Using AKARI/IRC spectroscopy, we discover a new sample of a heavily obscured AGN in LEDA~1712304 which shows a deep spectral absorption feature due to silicate dust.}
{We study the infrared (IR) spectral properties of circumnuclear silicate dust in LEDA~1712304.}
{We perform IR spectral fitting, considering silicate dust properties such as composition, porosity, size and crystallinity. Spectral energy distribution (SED) fitting is also performed to the flux densities in the UV to sub-millimeter range to investigate the global spectral properties.}
{The best-fit model indicates $0.1\,{\rm {\mu}m}$-sized porous amorphous olivine (${\rm Mg_{2x}Fe_{2-2x}SiO_4}$; $x=0.4$) with $4\%$ crystalline pyroxene. The optical depth is $\tau_{\rm sil}{\sim}2.3$, while the total IR luminosity and stellar mass are estimated to be $L_{\rm IR}=(5\pm1){\times}10^{10}\,L_{\odot}$ and $M_{\rm star}=(2.7\pm0.8){\times}10^{9}\,M_{\odot}$, respectively. In such low $L_{\rm IR}$ and $M_{\rm star}$ ranges, there are few galaxies which show that large ${\tau}_{\rm sil}$.}
{The silicate dust in the AGN torus of LEDA~1712304 has properties notably similar to those in other AGNs as a whole, but slightly different in the wing shape of the absorption profile. The porosity of the silicate dust suggests dust coagulation or processing in the circumnuclear environments, while the crystallinity suggests that the silicate dust is relatively fresh. } 
% 5 {} token are mandatory
 
 % \abstract

   %{}

\titlerunning{IR spectroscopy of the heavily obscured AGN LEDA~1712304}
\authorrunning{T. Tsuchikawa et al. }

   \maketitle

%
%-------------------------------------------------------------------
% and の前後が同等であるか
%文と文のつながり意識
%spectroscopy & spectroscopic data 20

\section{Introduction}
%The cosmic history of the star formation (SF) and active galactic nuclei (AGN) activities peaked at the redshift range of 1--3 \citep{Madou2014}.  
%In the epoch, the galaxies had extremely dusty environments. Thus the UV and optical emission from massive stars or the AGN accretion disk was mostly covered by interstellar dust or dusty torus which is suggested in the AGN unified model \citep{Antonucci1993,Netzer2015}, and then dust reemitted infrared (IR) lights. Accordingly, the IR dust emission reflects the SF and AGN activities deeply involved in the galaxy evolution. However, dust evolution such as size, chemical composition and crystallinity with the galaxy evolution is poorly understood. 
%z~1-2で銀河の活動度（SF&AGN）がピークを迎えたことが分かっており、こういった年代でこれらの活動を支えるダストがどう進化していったかしりたい。

%
%The dust-obscured galaxies (DOGs) are optically faint and mid-IR blight galaxies selected with the threshold of a ratio of the observed mid-IR 24$\,{\rm{\mu}m}$ to optical $\it R$ band flux, $F_{24\,{\rm{\mu}m}}/F_R > 1000$ \citep{Dey2008}. \citet{Dey2008} suggested that the high-z DOGs accounted for $60^{+40}_{-15}\,\%$ of $z\sim2$ ultraluminous infrared galaxies (ULIRGs; $L_{\rm IR}>10^{12}L_{\odot}$) and can be divided into SF-dominated DOGs and AGN-dominated DOGs on the basis of near-IR spectral energy distribution (SED). In a evolutionary scenario, DOGs are driven by a major merger, and naturally evolve from SF-dominated DOGs to AGN-dominated DOGs \citep{Narayanan2010} 
%その中でもdust-obscured galaxy は、深くダストに埋もれているため、可視光が強く減光を受け、赤外でとても明るい。

 Dust plays an important role in probing the activity of active galactic nuclei (AGN). The AGN unified model \citep[e.g.,][]{Antonucci1993,Urry1995} concludes that a supermassive black hole (SMBH) and its accretion disk are enshrouded by an optically thick dusty torus which has an axisymmetric structure to explain the difference in the optical line profiles. 
Heavily obscured AGNs which show almost no optical sign of AGN activities have been found by many observational studies \citep[e.g.,][]{Sanders1996,Lutz1998,Dey2008,Imanishi2008,Oyabu2011}. 
Due to the presence of thick dust layers around the AGN core, the UV and optical emission from the AGN accretion disk is strongly suppressed, and thus AGN identification on the basis of the optical line ratios encounters difficulty for the heavily obscured AGNs.  %a lot of galaxies are obscured at the redshift $z\sim2$, and declined from now on. so  
Infrared (IR) observations can identify AGNs without suffering severe dust extinction. 
For heavily obscured AGNs, the near- to mid-IR range is dominated by the hot dust emission heated by AGNs, where the spectra show power-law like continua. 
%Many ultraluminous infrared galaxies \citep[ULIRGs; ][]{Sanders1996} were suggested to host obscured AGNs \citep[e.g., ][]{Lutz1998}. 
While the presence of heavily obscured AGNs is suggested in many cases, the properties of the surrounding dust, such as size, chemical composition and crystallinity, are poorly understood.

%これらの天体はダストに深く埋もれているため、可視光で降着円盤からの放射が見えず、optical line ratioを使ったAGN診断が困難となる。

%In a evolutionary scenario, major mergers trigger a starburst phase and then AGN glow with accretion to a black hole and feedback.

%%Milky way GCはa shortage of carbon-rich stars (Roche&Aitken 1985;Whittet 1987)いる。

%AGNを持つdogの内AGN torusのdustはその中心の高エネルギーにより、変性等の影響を受けている可能性がある。Kaneda et al.では

%carbonaseous dust was degenerated 

%近中間赤外線の分光はSF/AGNの活動度とsilicate dustの性質の同時理解が可能であり有効。
% Silicate 10umSi-O, 18umO-Si-O　組成や構造を反映して変形、、、、%%%PAH,ice
%
The spectroscopy in the near- to mid-IR range which contains various dust spectral features is effective to understand the properties of dust. 
In particular, the profiles of the silicate features at around 9.7$\,{\rm{\mu}m}$ (Si--O stretching mode) and 18$\,{\rm{\mu}m}$ (O--Si--O bending mode) reflect the chemical composition, size distribution, crystallinity and structure of the silicate dust. 
The silicate features for nearby luminous IR galaxies (LIRGs) and ultra-luminous IR galaxies (ULIRGs) have been studied extensively by ${\it Spitzer}$/IRS \citep[e.g.,][]{Imanishi2009, Stierwalt2013}.
In particular, many U/LIRGs have deep silicate absorption features \citep[e.g.,][]{Roche1986,Spoon2007}, which tend to possess heavily obscured AGNs \citep[e.g.,][]{Imanishi2009} and are likely to be at the late to final stages of mergers \citep{Stierwalt2013}.    
On the other hand, the depths of the silicate features in AGNs depend on the AGN type; type-1 Seyferts show the features in either emission or weak absorption, while type-2 Seyferts show those only in absorption \citep{Hao2007}.
As a global trend, the number of galaxies which have deep silicate absorption features increases with the IR luminosities of their host galaxies \citep{Imanishi2009}.
\citet{Spoon2007} suggested that the difference in the silicate absorption depth depends on the circumnuclear distribution of the dusty torus.
\citet{Roche2015} reported that the silicate absorption profiles of NGC~4418 showed spectral variations depending on the slit aperture size, which suggests the presence of extended dust emission around the nucleus. 
Although the main compositions of the features are likely to be amorphous olivine, crystalline silicate substructures are also detected from heavily obscured ULIRG nuclei \citep{Spoon2006, Stierwalt2014}.

%The optical-to-silicate extinction ratios are $A_V/\tau_{9.7}<5.5$, which are smaller than those in our Galaxy, implying that type-2 AGNs have dust grains of larger sizes \citep{Lyu2014,Shao2017}

%%%because dust coagulation in the high-density environment in ULIRGs’ cores could increase typical dust grain size up to a few microns (Laor & Draine 1993;Maiolino et al. 2001a, 2001b; Imanishi 2001)
%main composition of interstellar silicate feature in our galaxy is amorphous olivine (Chiar Tielens). type-2 AGN NGC~1068 firstly detected (Rieke) . heavily obscured galaxy NGC4418 (Roche)

%Type-2 heavily obscured AGNs are known to exhibit the deep silicate absorption features \citep{}. 

%Moreover, polycyclic aromatic hydrovarbon (PAH) features at () can distinguish the activities in the galaxy. The PAH features are emitted dominantly from photo-dissociation regions (PDRs), and thus SF-dominated galaxies strongly emitted the features. However, in AGN-dominated galaxies, PAHs are distructed by X-ray plasma, and hot dust continuum are dominated in the wavelength range. 
 %the equivalent widths of PAH features get smaller in AGN-dominant galaxies. The steep near-IR continuum is also a probe of AGNs since dust surrounding AGN core can be heated up to $\sim500\,{\rm K}$.

%z~1-2の理解には重要。(Spoon2004)近傍では暗い銀河も受かる、

%銀河進化

%Spiter ではimanishi, shi ％％％AKARではImanishi

% extremely obscured AGN optically faint, mid infrared bright galaxies
%The AGN activity is especially more likely to affect dust, since dust can be exposed to higher energy environment.

In AKARI/IRC slit-less spectroscopic surveys, we serendipitously discover a galaxy which shows a notably deep silicate absorption feature but with a relatively low IR luminosity.
The galaxy is a local dust-obscured galaxy, LEDA~1712304 \citep[$z = 0.0645$;][]{Hwang2013}.
A galaxy classification of LEDA~1712304 based on the optical line ratios is ``undetermined'' \citep{Hwang2013}, which indicates that the dust extinction in the galaxy is too large to detect optical lines such as [OIII] 5007.  
On the other hand, the galaxy is expected to possess an AGN hidden in the galactic center based on the result of IR spectral energy distribution (SED) fitting \citep{Hwang2013}. 
In this paper, we study the spectral properties of silicate dust in LEDA~1712304, using the AKARI/IRC spectroscopic data as well as the photometric data in the UV to sub-millimeter range.
Throughout the paper, we adopt $\rm 289.7\,Mpc$ for the distance to the galaxy, assuming the cosmological parameters $H_0=70\,{\rm km\,s^{-1}\,Mpc^{-1}}$, ${\Omega}_{\Lambda}=0.7$ and ${\Omega}_m=0.3$.
% optical: non-seyfert, AGN fraction < 100%:according to SED fit, likely to have 
 
%Elliptical, coordinates, z, dog(Hwang) SDSS posess agn  %major axis:minor axis ~ 2:1
%merging group Tempel+, 2017

%In Sect.~2 and ~3, the observations and the results are reported. In Sect.~4, we discuss the properties of the dust-obscured galaxy LEDA1702304 based on the near- to mid-IR spectroscopy.  
%Throughout this paper, we adopt 289.7 Mpc for the distance to LEDA~1712304.
%------------------------------------------------------------------
\section{Observations}
\subsection{AKARI IR spectroscopy}
LEDA~1712304 was detected serendipitously within the field-of-view of slit-less spectroscopy at 2.5--12.5$\,{\rm{\mu}m}$ using InfraRed Camera \citep[IRC;][]{Onaka2007} onboard the AKARI satellite \citep{Murakami2007}, the mid-IR spectrum of which is among those in the catalog of \citep{Yamagishi2019}.
The primary target was IC~860 and the observation was carried out on July 28 2007 with the IRC spectroscopic mode of the astronomical observation template 04 (AOT04) in the framework of the mission program AGNUL (Evolution of ULIRGs and AGNs; PI: T. Nakagawa).
In this study, the near-IR spectrum is combined with the mid-IR spectra to create the 2.5--12.5$\,{\rm{\mu}m}$ of LEDA~1712304, where three spectroscopic modules were used; the NIR prism (NP) and the two MIR-S grisms (SG1, SG2) covering the wavelength ranges of 1.8--5.5$\,{\rm{\mu}m}$, 4.6--9.2$\,{\rm{\mu}m}$ and 7.2--13.4$\,{\rm{\mu}m}$, respectively.
The wavelength resolution, $\rm{R\,(=\lambda/\delta\lambda})$, is typically 19 at 3.5$\,{\rm{\mu}m}$, 53 at 6.6$\,{\rm{\mu}m}$ and 50 at 10.6$\,{\rm{\mu}m}$ for NP, SG1 and SG2, respectively \citep{Ohyama2007}. 
Since LEDA~1712304 was observed with the slit-less spectroscopy, the spectrum does not need to be corrected for the slit loss.
The spectra are processed with the data reduction package ``IRC Spectroscopy Toolkit Version 20181203''. 
For extraction of the spectrum, we applied the aperture sizes of 5 pixels (7.3$\arcsec$), 7 pixels (16.4$\arcsec$) and 7 pixels (16.4$\arcsec$) for NP, SG1 and SG2, respectively. 
The NP and SG spectra were calibrated on the basis of the AKARI 3.2 $\rm {\mu}m$ and 9 $\rm {\mu}m$ photometric flux densities, respectively. %下でのべる
We did not use the spectral regions of 2.0--2.5$\,{\rm{\mu}m}$ and 5.0--5.5$\,{\rm{\mu}m}$, which correspond to the edges of the NP spectral coverage, since systematic errors are relatively large in those regions. 
% Observation mode,(phase1,phase2 ,AGNUL,  slit-less) slit-less 不定性 Observation config, integration time, data reduction たまたま写ってた 積分時間

\subsection{Multi-wavelength photometric data}
In order to investigate the global spectral properties of LEDA~1712304, we conducted aperture photometry of the multi-wavelength image data taken with not only AKARI but also ${\it Spitzer}$ \citep{Fazio2004, Rieke2004} and {\it Herschel} \citep{Poglitsch2010, Griffin2010}. The ${\it Spitzer}$ and {\it Herschel} image data were retrieved from the ${\it Spitzer}$ Heritage Archive through the NASA/IPAC IR Science Archive (IRSA) and the ESA {\it Herschel} science archive, respectively. % obs program, pipeline 
We applied photometric apertures, $R_{\rm ap}$, in Table~\ref{table:flux}, and sky backgrounds were determined in the annuli of inner radii 1.2$R_{\rm ap}$ and outer radii 2$R_{\rm ap}$.
We also applied appropriate aperture corrections to the photometric results as mentioned in each of the handbooks.
Table~\ref{table:flux} summarizes the results of the aperture photometry. 
LEDA~1712304 was not detected in the {\it Herschel} 160$\,{\rm{\mu}m}$ band with the signal-to-noise ratio ($S/N$) > 3. 
\begin{table}
\caption{Summary of the aperture radii used for the photometry and the flux densities.}             % title of Table
\label{table:flux}      % is used to refer this table in the text
\centering                          % used for centering table
\begin{tabular}{l c c}        % centered columns (4 columns)
\hline\hline                 % inserts double horizXSontal lines
 Instrument&aperture radius / {[\arcsec]} \tablefootmark{a}&flux density / {[mJy]}\\    % table heading 
\hline                        % inserts single horizontal line
 GALEX NUV&--&$0.003\pm0.001$\tablefootmark{b}\\
 SDSS $\it{u}$&--&$0.043\pm0.005$\tablefootmark{b}\\
 SDSS $\it{g}$&--&$0.168\pm0.003$\tablefootmark{b}\\
 SDSS $\it{r}$&--&$0.347\pm0.003$\tablefootmark{b}\\
 SDSS $\it{i}$&--&$0.483\pm0.004$\tablefootmark{b}\\
 SDSS $\it{z}$&--&$0.59\pm0.02$\tablefootmark{b}\\
 SIRIUS $\it{J}$&12.7&$0.90\pm0.03$\\      % inserting body of the table
 SIRIUS $\it{H}$&12.0&$1.09\pm0.03$\\      % inserting body of the table
 SIRIUS $\it{K_s}$&12.3&$0.90\pm0.05$\\      % inserting body of the table
 IRC 3.2$\,{\rm{\mu}m}$&4.26&$2.57\pm0.06$\\      % inserting body of the table
 WISE 3.4$\,{\rm{\mu}m}$&--&$2.73\pm0.06$\tablefootmark{c}\\
 IRAC 3.6$\,{\rm{\mu}m}$&7.06&$4.16\pm0.08$\\      % inserting body of the table
 IRAC 4.5$\,{\rm{\mu}m}$&7.32&$8.65\pm0.18$\\      % inserting body of the table
 WISE 4.6$\,{\rm{\mu}m}$&--&$11.9\pm0.2$\tablefootmark{c}\\
 IRAC 5.8$\,{\rm{\mu}m}$&8.00&$28.8\pm0.6$\\      % inserting body of the table
 IRAC 8.0$\,{\rm{\mu}m}$&8.43&$45.3\pm0.9$\\      % inserting body of the table
 IRC 9$\,{\rm{\mu}m}$&--&$29\pm2$\tablefootmark{d}\\
 WISE 12$\,{\rm{\mu}m}$&--&$23.7\pm0.4$\tablefootmark{c}\\
 WISE 22$\,{\rm{\mu}m}$&--&$50\pm2$\tablefootmark{c}\\ 
 MIPS 24$\,{\rm{\mu}m}$&12.6&$50\pm2$\\      % inserting body of the table
 PACS 70$\,{\rm{\mu}m}$&11.5&$(1.8\pm0.2)\times10^2$\\      % inserting body of the table
 PACS 100$\,{\rm{\mu}m}$&14.3&$(3.0\pm0.3)\times10^2$\\      % inserting body of the table
 PACS 160$\,{\rm{\mu}m}$&24.0&<$\,140$\\      % inserting body of the table
 SPIRE 250$\,{\rm{\mu}m}$&37.4&$(5\pm1)\times10^1$\\      % inserting body of the table
 SPIRE 350$\,{\rm{\mu}m}$&50.9&$(4\pm1)\times10^1$\\      % inserting body of the table
\hline                                   %inserts single line

\end{tabular}
\tablefoot{
\tablefoottext{a}{The aperture radii $R_{\rm ap}$ correspond to 3, 10 and 5$\,{\times}\,R_{\rm FWHM}/2.35$ for SIRIUS, IRAC and the other image data, respectively, where $R_{\rm FWHM}$ is the full width at half maximum (FWHM) of the point spread function (PSF).}
\tablefoottext{b}{\citet{Hwang2013}.}
\tablefoottext{c}{\citet{Cutri2013}.}
\tablefoottext{d}{\citet{Yamagishi2019}.}
}% systematic error
\end{table}
%tableに書く？optical, UVはoptical spectroscopyで十分?
% \subsection{IRSF/SIRIUS near infrared photometry}
We also observed LEDA~1712304 with Simultaneous Infrared Imager for Unbiased Survey \citep[SIRIUS;][]{Nagashima1999,Nagayama2003} on Infrared Survey Facility (IRSF) in South African Astronomical Observatory. SIRIUS has a large field of view (7$\arcmin$ $\times$ 7$\arcmin$) and enables simultaneous imaging in the $\it{J}$, $\it{H}$ and $\it{K_s}$ bands. The observation was performed on June 6 2018, and the total integration time was 16.7 minutes. 
We performed the flux calibration on the basis of the 2MASS Point Source Catalog \citep{Cutri2003}.  
In addition, we also used the flux densities of LEDA~1712304 taken from the UV, optical and IR catalogs \citep{Hwang2013, Cutri2013, Yamagishi2019}.

\section{Results}
\subsection{IR spectral properties}
\begin{figure*}
   \centering
   \includegraphics[width=14cm]{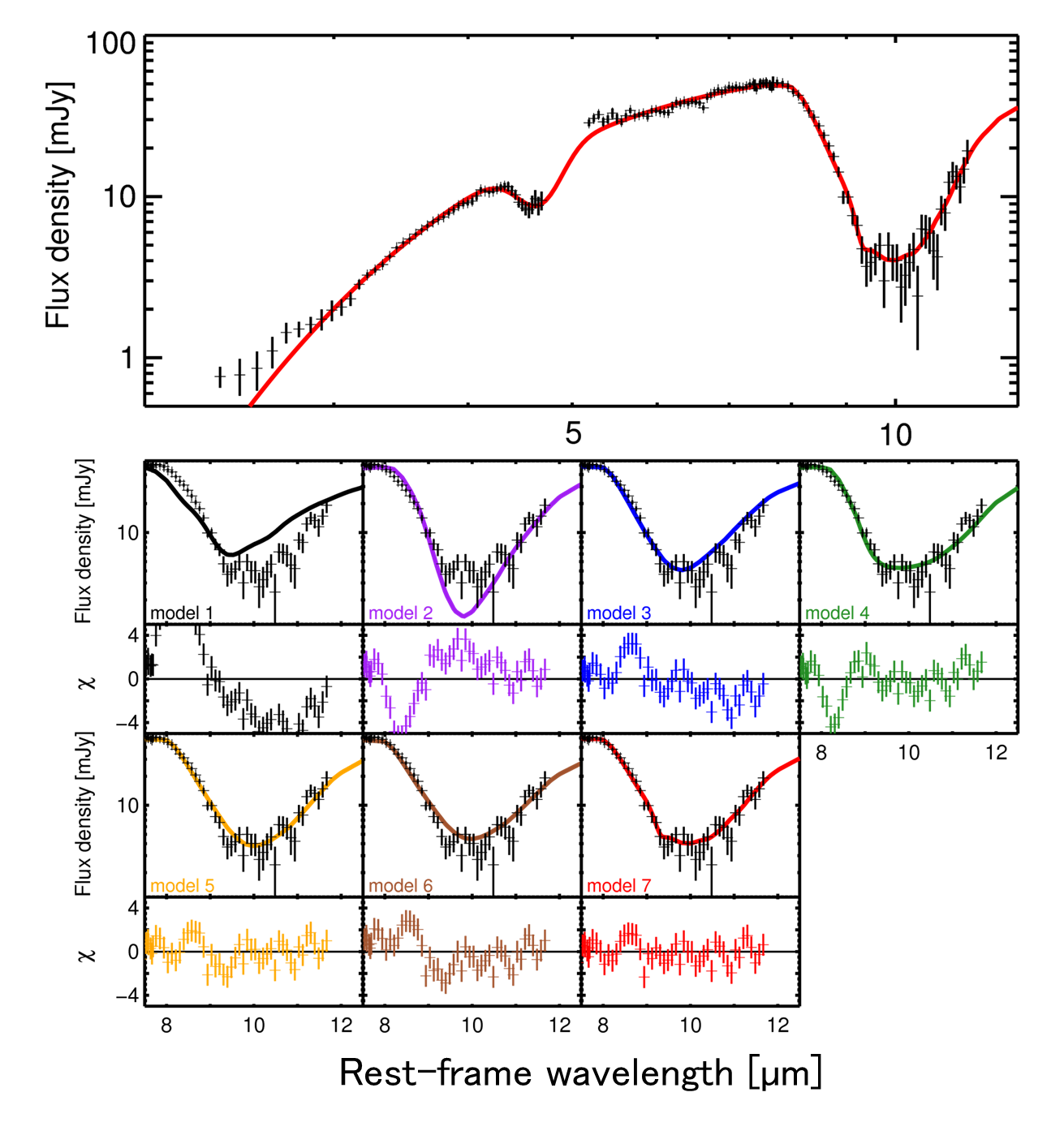}
   %\resizebox{\hsize}{!}{\includegraphics{fig1_5.png}}
   \caption{$\it Top$ $\it panel$: AKARI/IRC spectrum of LEDA~1712304 shown together with the best-fit model (model 7; red solid line). ${\it Bottom}$ $\it panel$: A closed-up view of the fitting results of the 10$\,{\rm{\mu}m}$ silicate absorption feature with various spectral models and the residuals normalized by the errors. Black, purple, blue, green, orange, brown and red solid lines show the astronomical silicate (model 1), amorphous olivine (${\rm Mg_{2x}Fe_{2-2x}SiO_4}$, $x=0.5$; model 2), amorphous olivine ($x=0.4$; model 3), a combination of the amorphous olivine ($x=0.5$) absorption and absorption-free emission component (model 4), the porous amorphous olivine ($x=0.4$; model 5), the $1\,{\rm {\mu}m}$-sized amorphous olivine ($x=0.4$; model 6) and a mixture of the porous amorphous olivine ($x=0.4$) and the crystalline pyroxene (model 7), respectively.}
              \label{spec}%
    \end{figure*}

Figure~\ref{spec} shows the AKARI/IRC 2.5--12.5$\,{\rm{\mu}m}$ spectrum of LEDA~1712304. 
We detect a deep, broad absorption feature due to silicate grains at around 10$\,{\rm{\mu}m}$. 
We also detect CO ro-vibrational absorption features at around 4.7$\,{\rm{\mu}m}$. %In addition, we identified a weak $\rm H_2O$ ice absorption feature around 3$\,{\rm{\mu}m}$. 
Any other spectral features, such as those due to $\rm CO_2$ ice at 4.27$\,{\rm{\mu}m}$, $\rm H_2O$ ice at 3.05 and 6.02$\,{\rm{\mu}m}$ and polycyclic aromatic hydrocarbon (PAH) features at 3.3, 6.2, 7.7 and 11.3$\,{\rm{\mu}m}$, are not detected significantly. 
The spectrum also shows a steep near- to mid-IR slope, which is likely to indicate that the continuum is dominated by thermal emission of hot dust heated by AGN. %\citep{kondo-san}. %6.2um H2Oice あるかも
According to the AKARI near-IR AGN diagnostics in \citet{Inami2018}, LEDA1712304 is indeed classified as AGN-dominated galaxy; for AGNs, the near-IR flux density ratio, $F_{\nu}(4.3\,{\rm {\mu}m})/F_{\nu}(2.8\,{\rm {\mu}m})$, and the equivalent width of the PAH $3.3\,{\rm {\mu}m}$ feature, $EW({\rm PAH}_{3.3\rm{{\mu}m} })$, are expected to be $>1.0$ and $< 0.06\,{\rm {\mu}m}$, respectively, while $F_{\nu}(4.3\,{\rm {\mu}m})/F_{\nu}(2.8\,{\rm {\mu}m})$ and the $3\sigma$ upper limit of $EW({\rm PAH}_{3.3\rm{{\mu}m} })$ are $7.5\pm0.9$ and $0.03\,{\rm {\mu}m}$, respectively, for LEDA~1712304, which is estimated by a Gaussian function with a local power-law continuum.

We evaluate the spectral properties through model fitting to the 2.5--12.5$\,{\rm{\mu}m}$ spectrum. %温度勾配急
In previous studies with similar AGN modeling, for example, \citet{Sani2008} used a steep power-law continuum to reproduce a 3--$5\,{\rm {\mu}m}$ spectrum, while \citet{Armus2007} used a hot-temperature blackbody with a cooler amorphous silicate absorption feature for a spectral range similar to ours. 
Referring to the latter, in reproducing the hot dust continuum and the deep spectral absorption feature, we assume a two-layer configuration for a dusty AGN torus, which is composed of a hot emitting inner layer and a warm absorbing outer layer. The hot emitting inner layer represents an optically thick hot dust layer which enshrouds the AGN core, while the warm absorbing outer layer represents a warm dust layer enveloping the hot inner layer. 
Hence the fitting function is described as
   \begin{equation}
    I_{\nu} = C_{1}{\rm exp}(-N_{\rm dust}{\pi}a^2Q_{{\rm abs}, {\nu}} - \tau_{\rm CO})B_{\nu}(T_{\rm h}),
   \label{eq:spec_model_pre}
   \end{equation}
where $N_{\rm dust}$, $a$, $Q_{{\rm abs}, {\nu}}$, ${\tau}_{\rm CO}$ and $B_{\nu}(T_{\rm h})$ are the column density, the size, the absorption efficiency of silicate dust in the outer layer, the optical depth of the CO absorption feature and the Planck function with the temperature, $T_{\rm h}$, of the hot dust emission from the optically thick inner layer, respectively. 
The dust size, $a$, is fixed at 0.1$\,{\rm {\mu}m}$, unless otherwise stated.
For the CO gas absorption profile, we adopted a Gaussian function, the center of which was fixed at 4.67$\,{\rm{\mu}m}$.

As a first step, we assume the standard astronomical silicate \citep{Draine2003b} tentatively for the composition of dust in the absorbing outer layer. %\footnote{This model was obtained in the Bruce T. Draine's Home Page ($\rm ~ftp://ftp.astro.princeton.edu/draine/dust/diel/callindex.out\_silD03$).  } 
The fitted model is shown by a black line in the bottom panel of Fig.~\ref{spec}, while the fitting result is summarized in Table~\ref{table:spec_para}.
We find that the standard astronomical silicate model does not reproduce the observed silicate absorption feature at all; not only the observed peak but also the feature profile is considerably different from those predicted by the model, as seen in the bottom panel of Fig.~\ref{spec}.
In principle, the peak of the model feature can be shifted toward longer wavelengths by changing either the size or porosity of silicate dust \citep{Laor1993,Li2008}. 
However we find that the feature profile of larger-size or porous astronomical silicate model then becomes too wide to reproduce the observed spectrum.

It is known that silicate features are fitted well with the composition of amorphous olivine (${\rm Mg_{2x}Fe_{2-2x}SiO_4}$; $x=0.5$) rather than the astronomical silicate for heavily obscured AGNs and Galactic center sources \citep{Spoon2006, Kemper2004}.
Hence we test two kinds of models of amorphous olivine \citep[${\rm Mg_{2x}Fe_{2-2x}SiO_4}$; $x=0.5$ and 0.4 for models 2 and 3, respectively; ][]{Dorschner1995}. 
The fitting results are listed in Table~\ref{table:spec_para}, which show that model 3 fits the spectrum better than model 2.
The best-fit models of amorphous olivine ($x=0.5$ and $x=0.4$) are also shown in the bottom panel of Fig.~\ref{spec}, where we find that the feature profile is still not reproduced well by either model, although the peak wavelengths show better fits to the spectrum as compared to model 1. %近づいただけでは？
%%%%%%%%%%%%%%%%%%%%%%%%%%%%%%%%%%%%%%%%%%%%%%%%%%%%%%%%%%%

%%%%%%%%%%%%%%%%%%%%%%%%%%%%%%%%%%%%%%%%%%%%%%%%%%%%%%%%%%%
Model 2 does not fit the spectrum very well at the bottom of the profile, while model 3 reproduces the bottom of the profile fairly well.
In order to improve the former discrepancy, we add an absorption-free hot dust emission component in the model, which raises the floor level of the feature, as pointed out by \citet{Spoon2004}. %, introducing a clumpy torus to reproduce the spectrum at the bottom of the silicate absorption feature. %\citep{Nenkova2002}. 
Thus the fitting function of equation (1) is modified as follows:
   \begin{equation}
    I_{\nu} = C_{1}{\rm exp}(-N_{\rm dust}{\pi}a^2Q_{{\rm abs}, {\nu}} - \tau_{\rm CO})B_{\nu}(T_{\rm h}) + C_{2}B_{\nu}(T_{\rm h}).
   \label{eq:spec_model}
   \end{equation}
%$Q_{{\rm abs}, {\nu}}$ was calculated from homogeneous sphere Mie theory \citep{BH1983}. 7
%Therefore the number of free parameter of the spectral fitting is 4.  
The result of the spectral fitting is shown by a green line in the bottom panel of Fig.~\ref{spec} (model 4) and also in Table~\ref{table:spec_para}, where we find that model 4 improves the fit around the bottom of the profile and yet the fitting is still not acceptable on the basis of the ${\chi}^2$ statistics.
In particular, model 4 fails to reproduce the wing of the profile on the shorter wavelength side.

%porous dust is Li2007 Min
On the other hand, since model 3 predicts a feature profile narrower than the observed absorption feature, we change the porosity and size of amorphous olivine ($x=0.4$) to widen the feature profile of model 3. 
%We can not apply them to the amorphous olivine ($x=0.5$) model, since the size and porosity of dust model widen silicate feature only toward longer wavelengths.
We calculate the absorption efficiency of porous dust, averaging dielectric functions according to the Maxwell Garnett theory \citep{BH1983}.  %, while we calculated the silicate dust models described above assuming compact homogeneous grains. 
Assuming that the porosity of amorphous olivine ($x=0.4$) is 30\%, which is a volume fraction of the vacuum in a dust grain, we improve the fit to the silicate feature significantly (model 5; see the orange line in Fig.~\ref{spec}), except for the residual seen at around 9.3$\,{\rm{\mu}m}$. 
We also test a large-size silicate model (model 6), assuming $1\,{\rm {\mu}m}$-sized amorphous olivine ($x=0.4$), instead of the porous amorphous olivine.
We however find that the fitting is only slightly improved from model 3 to model 6, and thus we conclude that the higher porosity is more favorable than the larger sizes for silicate grains.

Even with model 5, we find that there are systematic residuals at around 9.3$\,\rm {\mu}m$, which degrade the fitting near the bottom of the absorption profile significantly.
Therefore we further add a crystalline pyroxene component \citep{Jager1994} to the porous amorphous olivine component.
Hence model 7 is a mixture of the porous amorphous olivine ($x=0.4$) and crystalline pyroxene, which results in the best fitting among the silicate models we tested, as shown in Table~\ref{table:spec_para}.

\begin{table*}
\caption{Summary of the results of the IR spectral fitting with various silicate dust models.}             % title of Table
\label{table:spec_para}      % is used to refer this table in the text
\centering                          % used for centering table
\begin{tabular}{c p{5.8cm} c c c c}        % centered columns (4 columns)
\hline\hline             % inserts double horizXSontal lines
   model & composition & $\chi^2/{\rm dof}$\tablefootmark{a} &  porosity & $T_{\rm h}$ [K] & $N_{\rm dust}$ [$10^{11} \,\rm cm^{-2}$] \\ % table heading 
\hline                      % inserts single horizontal line
  1 & astronomical silicate  & 24.5 & -- & $486\pm2$ &  $0.573\pm0.007$ \\ 
  2 & amorphous olivine ($x=0.5$) & 6.4  & -- &  $495\pm1$ & $1.02\pm0.02$ \\
  3 & amorphous olivine ($x=0.4$) & 3.4 & -- &  $500\pm1$ & $0.95\pm0.01$ \\ 
  4 & amorphous olivine ($x=0.5$) & 3.1 & -- &  $494\pm1$ & $1.36\pm0.04$ \\ 
     &   + absorption-free continuum & & & &\\	
  5 & porous amorphous olivine ($x=0.4$) & 1.6 & 0.3 &  $502\pm1$ & $1.06\pm0.01$ \\
  6 & $1\,{\rm {\mu}m}$-sized amorphous olivine ($x=0.4$) & 2.7 & -- &  $552\pm1$ & $0.080\pm0.001$ \\ 
  7 & 96\% porous amorphous olivine ($x=0.4$)  & 1.2 & 0.3 & $493\pm1$ & $1.05\pm0.01$ \\ 
    &  + 4\% crystalline pyroxene (96:4) & & & & \\
 \hline                                   %inserts single line
\end{tabular}
\tablefoot{
\tablefoottext{a}{$\chi^2$ is calculated for the rest-frame spectral range of 8--12$\,{\rm{\mu}m}$. The degree of freedom (dof) is 35.}
%\tablefoottext{b}{\citet{Cutri2013} : ALLWISE}
%\tablefoottext{c}{Yamagishi et al. in prep.}
}
\end{table*}

%free parameter equation, model picture, Qabs, chi2, no PAH co2 ...,  実際のfittingではCOは外してfittingする？

\subsection{Multi-wavelength photometric properties\label{sect:sed}}

\begin{figure}
   \centering
   \resizebox{\hsize}{!}{\includegraphics{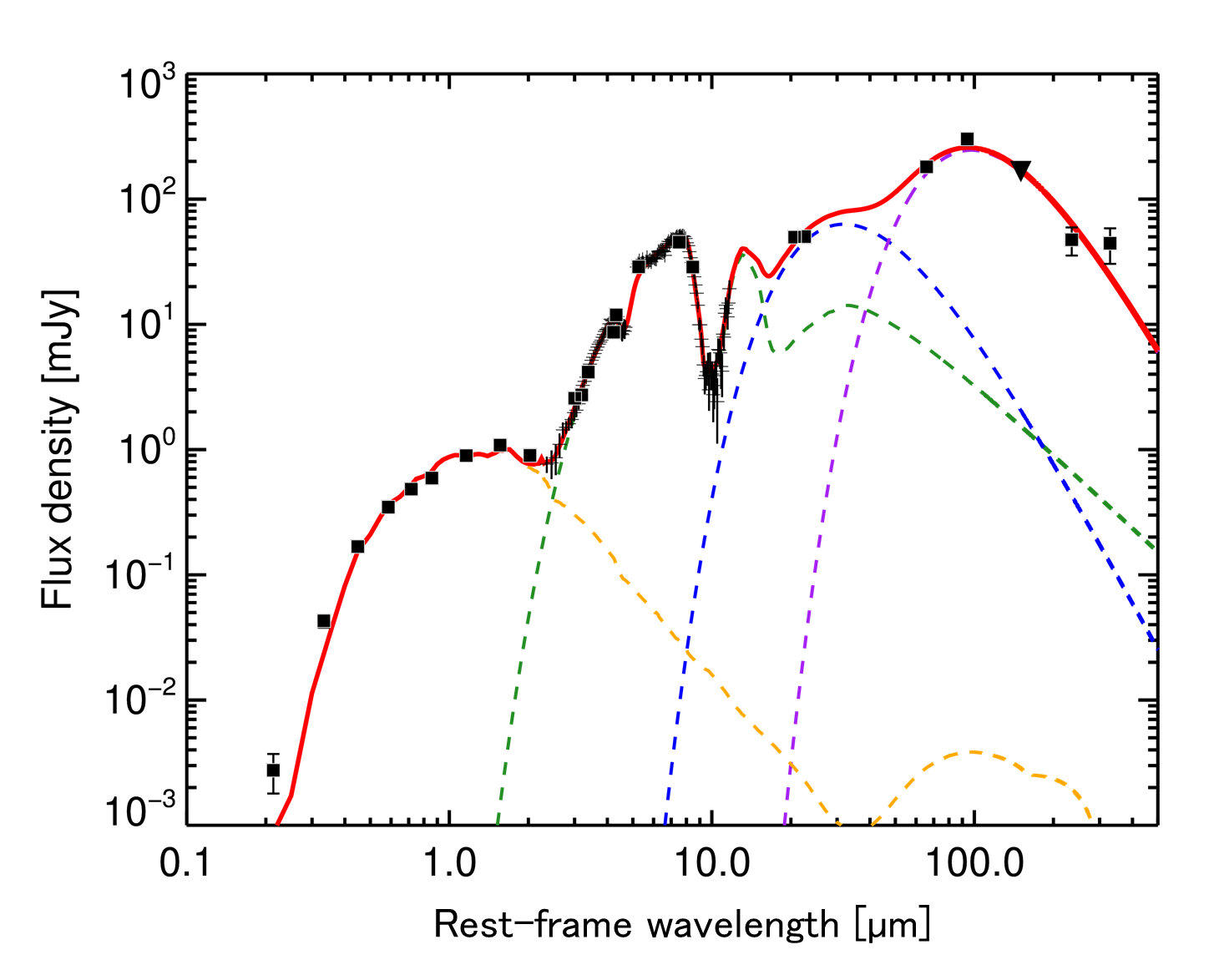}}
   \caption{Spectral energy distribution of LEDA~1712304. Filled squares correspond to the observed flux densities, while a downward triangle represents the $3\sigma$ upper limit. The AKARI/IRC spectral data at 2.5--12.5$\,{\rm{\mu}m}$ are also included in the fitting. A red solid line shows the best-fit model, where the stellar, hot dust, warm dust and cold dust components are shown by yellow, green, blue and purple dashed lines, respectively.}  %國生さん直し
              \label{sed}%
    \end{figure}

Figure~\ref{sed} shows the SED of LEDA~1712304, which is created with the flux densities listed in Table~\ref{table:flux}.
We perform spectral decomposition using the four SED components of stellar, hot dust, warm dust and cold dust emission. 
We confirm that the photometric flux densities at 2.5--12.5$\,{\rm{\mu}m}$ except that of the WISE 12$\,{\rm{\mu}m}$ band which contains the silicate feature are consistent with the AKARI/IRC spectrum.
We used the AKARI/IRC spectrum as well as the near- to mid-IR flux densities except the WISE 12$\,{\rm{\mu}m}$ flux density for the SED fitting. 
As can be seen in Fig.~\ref{sed}, the SED exhibits that the UV to near-IR continuum is dominated by notably red stellar emission. 
Hence, in the SED fitting, we apply a stellar continuum model for an elliptical galaxy \citep[13 Gyr;][]{Silva1998} and also take into account the reddening effect, for which the dust extinction law derived in \citet{Calzetti2000} is adopted. % calzetti は star domonant だけどいい？
Since the near- to mid-IR range is dominated by hot dust emission, we adopt the best-fit silicate model derived from the spectral fitting (model 7) for the hot dust component. %
 In the far-IR range, we find that double-temperature dust emission reproduces the SED fairly well. %These two dust components cannot explain the flux densities at 22 and $24\,{\rm {\mu}m}$. Thus we introduce a warm dust component. 
For the warm and cold dust components, we assume modified blackbody emission with the emissivity power-law index $\beta=2$. 

Based on the result of the SED fitting, we derive the stellar mass, $M_{\rm star}$, the IR luminosity, $L_{\rm IR}$, and the dust mass, $M_{\rm dust}$, which are listed in Table~\ref{table:sed}. 
$M_{\rm star}$ is estimated by assuming the stellar-mass-to-luminosity ratio, $M_{\rm star}/L_{K_s}\sim 1$ \citep{Cole2001}.   %, which result in $(2.7\pm0.8)\times10^{9}\,M_{\odot}$. %%%%%% Hwang et alだと5倍
%On the other hand, the stellar mass on SDSS DR15 \citep{Aguado2019} which is estimated by Principal Component Analysis \citep{Chen2012} is (1.3--1.9)$\times10^{10}\,M_{\odot}$.
%The inconsistency of the stellar masses is likely attributed to an uncertainty of the stellar population synthesis models we assume.
$L_{\rm IR}$ is calculated to be $(5\pm1)\times10^{10}\,L_{\odot}$ by integrating the best-fit model spectrum over the wavelength range of 8--1000$\,{\rm {\mu}m}$, which is consistent with the lower limit estimated by \citet[][$L_{\rm IR} > 3.42 \times 10^{10}L_\sun$]{Hwang2013}. 
The warm and cold dust emission is likely to be optically thin in the far-IR range, and we can derive $M_{\rm dust}$ as 
\begin{equation}
    M_{\rm dust} = \frac{F_{\nu}D^2}{B_{\nu}(T)\kappa_{\nu}},
   \label{eq:dust_mass}
\end{equation}
where $F_{\nu}$, $D$, and $\kappa_{\nu}$ are the flux density, the distance to the galaxy and the mass absorption coefficient, respectively.
$\kappa_{\nu}$ is assumed to be proportional to ${\nu}^2$, and we adopt $\kappa_{140\,{\rm {\mu}m}}=13.9\,{\rm cm^2/g}$ \citep{Draine2003}.
We also calculate the optical depth of the silicate feature, ${\tau}_{\rm sil}$, following the method defined by \citet{Imanishi2007}, where an absorption-free continuum is assumed to be the power-law function determined from the flux densities at 7.1 and 14.2$\,{\rm {\mu}m}$.
As a result, ${\tau}_{\rm sil}$ is estimated to be $2.3\pm0.2$ at 9.8$\,{\rm {\mu}m}$, the wavelength of the bottom of the absorption feature.

We also performed the SED fitting of the heavily obscured AGN, using the bayesian-based SED fitting code CIGALE \citep{Burgarella2005,Noll2009,Boquien2019} which includes energy balance between the dust absorption of the stellar/AGN emission and the dust re-emission in the mid/far-IR, similarly to the study by \citet{Ciesla2015}. As a result, $M_{\rm star}$ and $L_{\rm IR}$, are estimated to be $8{\times}10^{9}\,M_{\odot}$ and $6{\times}10^{10}\,L_{\odot}$, respectively. Comparing the result in Table~\ref{table:sed}, we find that $M_{\rm star}$ increases while $L_{\rm IR}$ does not change. The increase in $M_{\rm star}$, however, does not change our conclusion below. %re chi2 = 38.4 (photometric), 4.5 (all)

\begin{table*}
\caption{Luminosities and masses of LEDA~1712304 obtained with the SED fitting.}             % title of Table
\label{table:sed}      % is used to refer this table in the text
\newlength{\myheight}
\setlength{\myheight}{0.5cm}
\centering                          % used for centering table
\begin{tabular}{c c c c}        % centered columns (4 columns)
\hline\hline                 % inserts double horizXSontal lines
 $L_{\rm IR}$ [$L_{\odot}$]  & $L_{\rm hot}$ [$L_{\odot}$] & $L_{\rm warm}$ [$L_{\odot}$] &  $L_{\rm cold}$ [$L_{\odot}$]   \\
\hline 
 $(5\pm1)\times10^{10}$  & ($4.20\pm0.06)\times10^{10}$  &  ($1.8\pm0.8)\times10^{10}$ &  $(2.3\pm0.6)\times10^{10}$ \\
\hline    
$M_{\rm star}$[$M_{\odot}$] \tablefootmark{a} & $M_{\rm dust, w}$ [$M_{\odot}$]  &  $M_{\rm dust, c}$ [$M_{\odot}$]  &  \\
\hline
 $(2.7\pm0.8)\times10^{9}$ & $(1.1\pm0.5)\times10^{4}$ & $(1.1\pm0.3)\times10^{7}$\\     
%
%   	$M_{\rm star}$[$M_{\odot}$] \tablefootmark{a} &  $(2.6\pm0.8)\times10^{9}$ \\
%	 $L_{\rm IR}$ [$L_{\odot}$]  & $(4.6\pm0.9)\times10^{10}$ \\
% 	$L_{\rm hot}$ [$L_{\odot}$] & ($4.15\pm0.06)\times10^{10}$  \\
%	$L_{\rm warm}$ [$L_{\odot}$] &  ($1.6\pm0.6)\times10^{10}$ \\
% 	$L_{\rm cold}$ [$L_{\odot}$] &  ($1.7\pm0.6)\times10^{10}$  \\
%	$M_{\rm dust, w}$ [$M_{\odot}$]  & $(5\pm2)\times10^{3}$\\
%	  $M_{\rm dust, c}$ [$M_{\odot}$] & $(6\pm2)\times10^{6}$  \\ % table heading 

\hline                                   %inserts single line
\end{tabular}
\tablefoot{
\tablefoottext{a}{The uncertainty corresponds to that in the adopted initial mass function \citep{Cole2001}.}
%\tablefoottext{b}{IR luminosity.}
%\tablefoottext{b,c,d}{The luminosity of the hot, warm and cold dust component.}
%\tablefoottext{c}{Warm dust luminosity.}
%\tablefoottext{d}{Cold dust luminosity.}
%\tablefoottext{e}{Cold dust mass.}
}
\end{table*}

\section{Discussion} 
\subsection{Comparison with other AGNs}

\begin{figure}
   \centering
   \resizebox{\hsize}{!}{\includegraphics{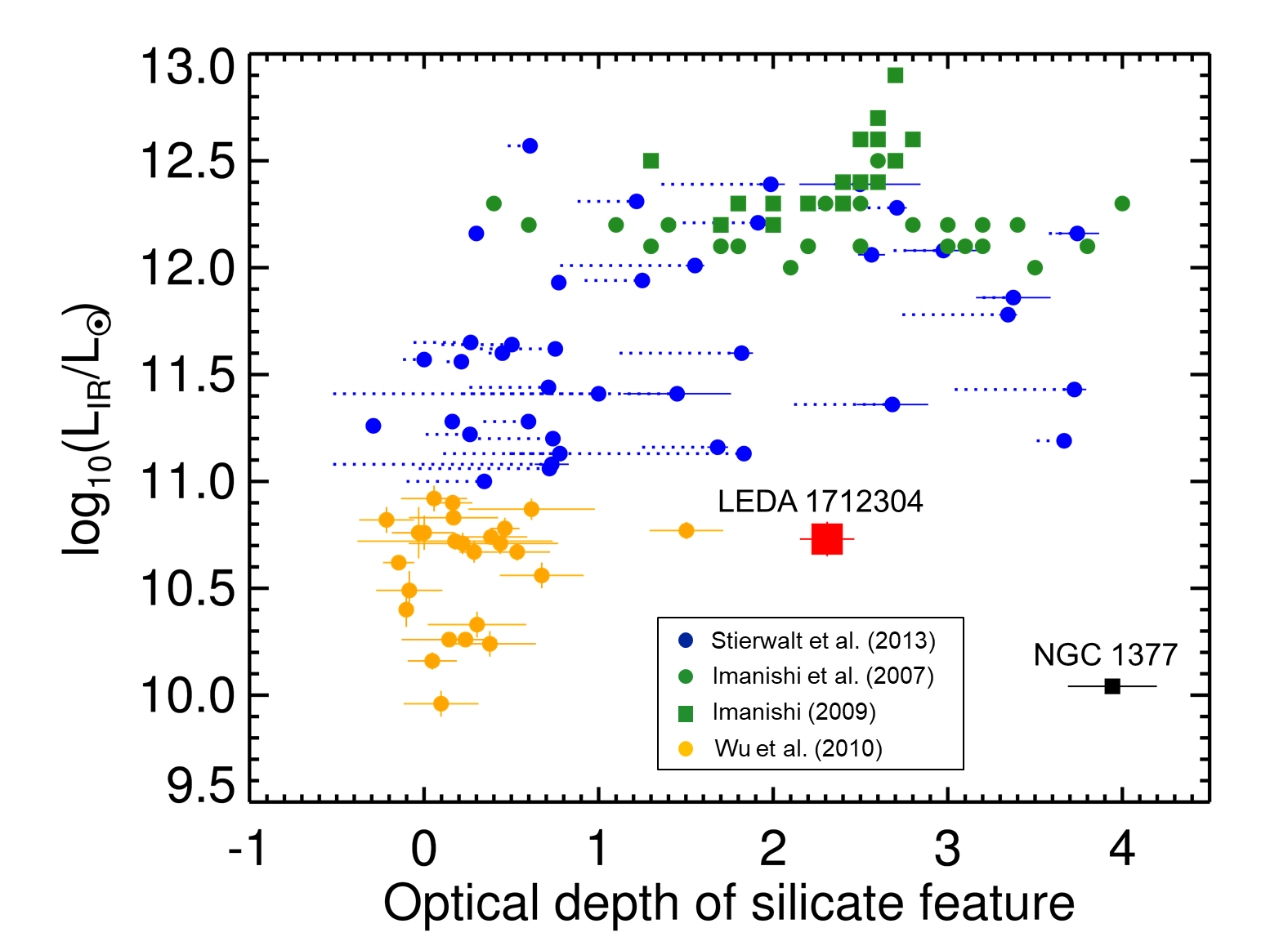}}
   \caption{Relation between the IR luminosity and the optical depth of the silicate feature for LEDA~1712304 and other AGN samples \citep{Stierwalt2013, Imanishi2007, Imanishi2009, Roussel2006, Wu2010}. Red and black squares represent LEDA~1712304 and NGC~1377 \citep{Roussel2006}, respectively. Blue, green, orange circles and green squares represent the AGN samples of \citet{Stierwalt2013}, \citet{Imanishi2007}, \citet{Wu2010} and \citet{Imanishi2009}, respectively. We use the IR luminosities obtained in each reference paper.
We estimate the optical depth of the silicate feature for LEDA 1712304 and the AGN samples of \citet{Stierwalt2013}, \citet{Wu2010} and \citet{Roussel2006} by ourselves, using the method defined by \citet{Imanishi2007}, while we take the values in each reference paper for the other AGNs. 
We also show the differences between our estimates and those by \citet{Stierwalt2013} with the blue dotted horizontal lines. } 
              \label{tau_lir}%
    \end{figure}
We detect a deep ($\tau_{\rm sil} \sim 2.3$) absorption feature due to silicate grains at around 10$\,{\rm{\mu}m}$.  
We compare the optical depth of the silicate feature in LEDA~1712304 with those in other AGNs in a wide range of IR luminosities ($10^{10}L_{\odot}<L_{\rm IR}<10^{13}L_{\odot}$). 
The spectra of the AGNs to be compared are taken from those of the IR galaxy samples observed by ${\it Spitzer}$/IRS \citep{Stierwalt2013, Imanishi2007, Imanishi2009, Roussel2006} with the threshold that the equivalent width of the PAH $6.2\,{\rm {\mu}m}$ feature is smaller than 0.27$\,{\rm {\mu}m}$ \citep{Stierwalt2013}.  % which trace AGN hot dust continua or PAH destruction by AGN activity.
In addition, we also take AGN samples with low IR luminosities ($L_{\rm IR}<10^{11}L_{\odot}$) from \citet{Wu2010}.
Figure~\ref{tau_lir} shows the relation between the IR luminosity and the optical depth of the silicate feature for LEDA~1712304 and the AGN samples.
We estimate the optical depths of the AGN samples of \citet{Stierwalt2013}, \citet{Wu2010} and \citet{Roussel2006} by ourselves in the same manner as performed for LEDA~1712304, with the method defined by \citet{Imanishi2007}, the spectra of which are obtained from the NASA/IPAC IR Science Archive (IRSA).
On the other hand, we adopt the values given in each reference paper for the other AGNs.
The blue dotted lines in Fig.~\ref{tau_lir} show the differences between our estimates and those by \citet{Stierwalt2013}, from which we confirm that the differences between the different methods are not as large as to change a global relation.
Fig.~\ref{tau_lir} exhibits that galaxies with low IR luminosities ($L_{\rm IR}<10^{11}\,L_{\odot}$) show significantly shallow silicate absorption features as already pointed out by \citet{Stierwalt2013}.
%\citet{Stierwalt2013} studied mid-IR properties of 202 nearby U/LIRGs at distances of $15\,{\rm Mpc}-400\,{\rm Mpc}$, which corresponds to the Great Observatories ALL-sky LIRG Survey (GOALS) sample \citep{Armus2009}, in addition to some star-forming galaxies with lower IR luminosities.
%They show that the galaxies without deep silicate absorptions $\tau_{\rm sil} > 1.5$ have a negative correaltion between the IR luminosities and the absorption depths, while those with deep absorptions $\tau_{\rm sil} > 1.5$ are widely scattered in the luminosity range of the GOALS sample, which is also shown as blue circle in Fig.~\ref{tau_lir}.
\citet{Imanishi2009} suggested that the number of heavily obscured AGNs which have deep silicate absorption features (${\tau}_{\rm sil}>2$) increases with the IR luminosities of the host galaxies.
Therefore LEDA~1712304 may be a rare galaxy from the aspect of having both deep absorption feature $\tau_{\rm sil} \sim 2.3$ and low IR luminosity $(5\pm1)\times10^{10}\,L_{\rm \odot}$.
Indeed, such galaxies have hardly been observed; an exception is NGC~1377 \citep{Roussel2006}, as can be seen in Fig.~\ref{tau_lir}.
NGC~1377 is a lenticular galaxy \citep{DeVau1991}, the stellar mass of which is $10^{9.3\pm0.1}\,M_{\odot}$ \citep{Skibba2011}.
The IR spectrum of NGC~1377 shows a featureless continuum except the silicate feature due to circumnuclear dust \citep{Imanishi2006,Roussel2006}.
%Hence NGC~1377 is similar to LEDA~1712304 in above properties, and therefore is considered to be in an evolutionary stage similar to LEDA~1712304.   %NGC１３７７も同じような進化phaseにいるrareな天体の一つと考えられる。

The low IR luminosity reflects a relatively low star formation rate (SFR) in LEDA~1712304. 
We applied the cold dust luminosity $L_{\rm cold}$ in Table~\ref{table:sed} instead of $L_{\rm IR}$ to estimate the SFR, since $L_{\rm IR}$ is likely to be contaminated with the AGN activity. 
The resultant SFR is $4\,M_\odot$/yr, based on the $L_{\rm IR}$--SFR relation in \citet{Kennicutt2012}. %もともとないとかは？
Hence the star formation is not active in the host galaxy, which is also expected from non-detection of the PAH features as well as the stellar continuum model of an elliptical galaxy for the SED fitting. 
\citet{Kauffmann2003} reported that the stellar masses of AGN host galaxies range from $10^{9.5}\,M_{\odot}$ to $10^{12}\,M_{\odot}$ based on the SDSS survey, whereas LEDA~1712304 has $M_{\rm star}=(2.7\pm0.8)\times10^{9}\,M_{\odot}$ (or $8{\times}10^9\,M_{\odot}$ from the CIGALE fitting; see Section~\ref{sect:sed}). 
Therefore LEDA~1712304 belongs to the population of a considerably low mass class which harbors an AGN. 
Thus the host galaxy is expected to be an early-type galaxy of low mass classes and yet possesses a heavily observed AGN; LEDA~1712304 is likely to belong to a population missed in the previous observations, and aforementioned NGC~1377 may be in an evolutionary stage similar to LEDA~1712304.

\begin{figure*}
   \centering
   \includegraphics[width=14cm]{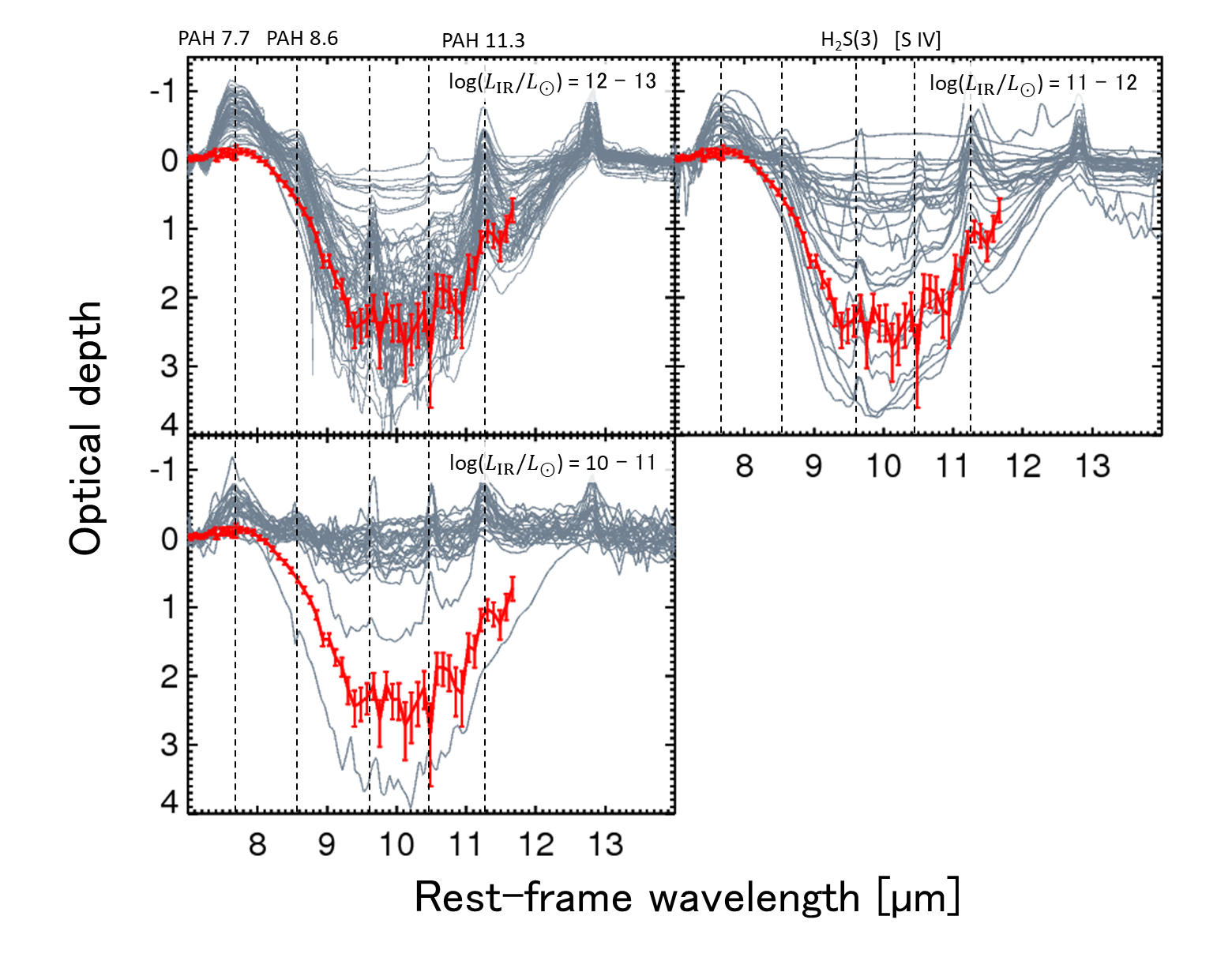}
   %\resizebox{\hsize}{!}{\includegraphics{fig4_4.png}}
   \caption{Optical depth profiles of the silicate features for LEDA~1712304 (red) and the other AGN samples \citep[grey;][]{Stierwalt2013, Imanishi2007, Imanishi2009, Roussel2006, Wu2010}. 
			$\it Top$ $\it right$, $\it top$ $\it left$ and $\it bottom$ $\it panels$ show the optical depth profiles of the AGN samples in the IR luminosity ranges of $10^{12}$--$10^{13}\,L_{\odot}$,  $10^{11}$--$10^{12}\,L_{\odot}$ and $10^{10}$--$10^{11}\,L_{\odot}$, respectively.
			Each optical depth profile is obtained by the absorption-free power-law continuum determined from the flux densities at $7.1\,{\rm{\mu}m}$ and $14.2\,{\rm{\mu}m}$ \citep{Imanishi2007}.
			The dashed lines show the positions of PAH 7.7, 8.6 and 11.3$\,\rm{\mu}m$, $\rm H_2~S(3)$ 9.66$\,\rm{\mu}m$ and [\ion{S}{IV}] 10.5$\,\rm{\mu}m$.} 
              \label{profile_plot}%
    \end{figure*}
\begin{figure*}
   \centering
   \includegraphics[width=14cm]{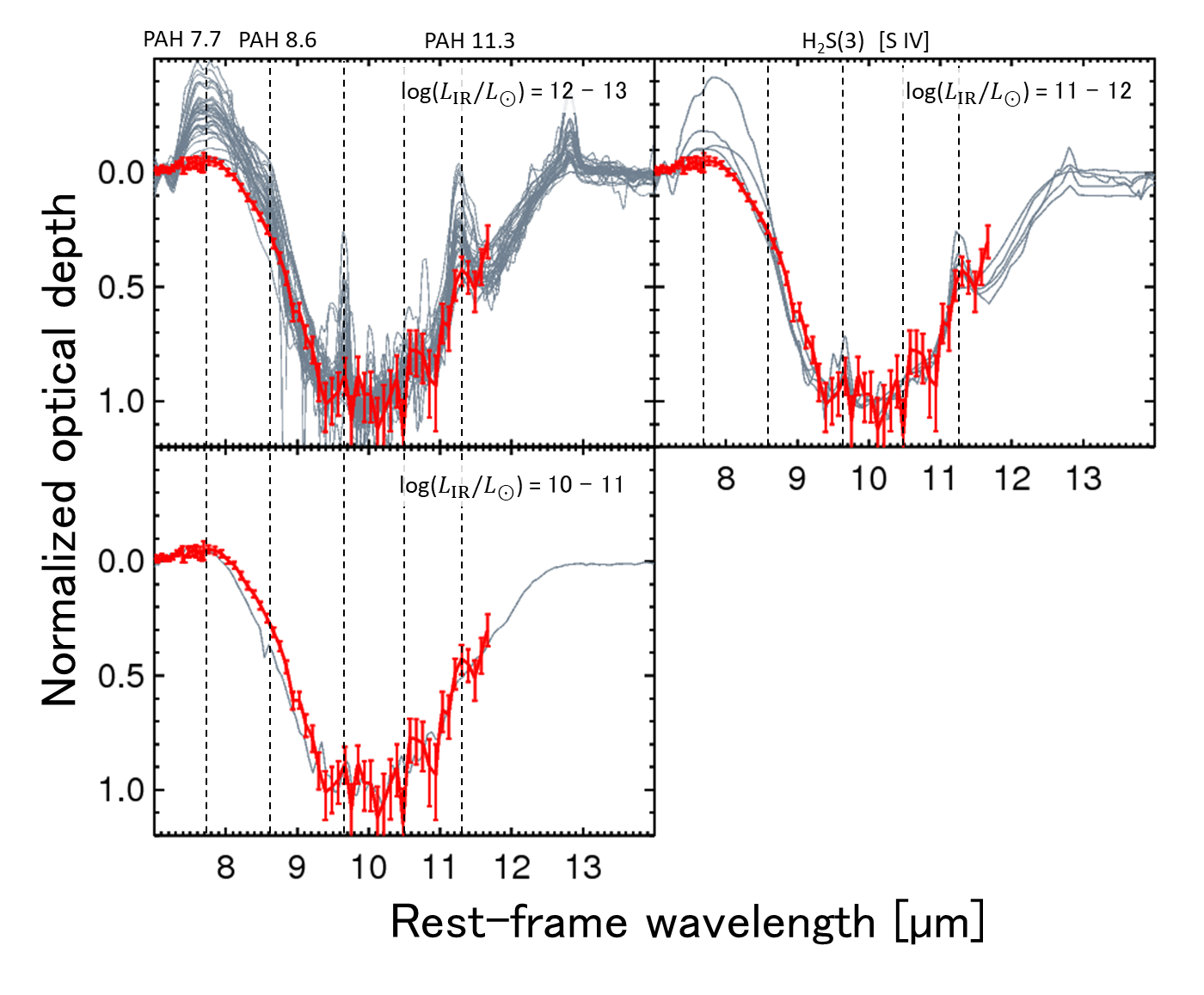}
   %\resizebox{\hsize}{!}{\includegraphics{fig5_8.png}}
   \caption{Normalized optical depth profiles of the silicate features for LEDA~1712304 (red) and the other AGN samples \citep[grey;][]{Stierwalt2013, Imanishi2007, Imanishi2009, Roussel2006} which have deep silicate absorption features (${\tau}_{\rm sil}>2$). 
			The normalization is performed by the mean optical depths at the wavelength range of 9.8--10.3$\,{\rm {\mu}m}$ which include no strong lines.
			The dashed lines show the positions of PAH 7.7, 8.6 and 11.3$\,\rm{\mu}m$, $\rm H_2~S(3)$ 9.66$\,\rm{\mu}m$ and [\ion{S}{IV}] 10.5$\,\rm{\mu}m$.} 
              \label{smoothed}%
    \end{figure*}

In Figure~\ref{profile_plot}, we compare the silicate feature profile of LEDA~1712304 with those of the other AGNs, classifying them with $L_{\rm IR}$. 
Fig.~\ref{profile_plot} shows that more AGN samples in a lower $L_{\rm IR}$ class tend to have shallower silicate absorption features (${\tau}_{\rm sil}<1$), which is consistent with the trend in Fig.~\ref{tau_lir}.
Fig.~\ref{profile_plot} also shows that the spectra of many AGN samples exhibit strong emission features, such as PAH $7.7\,{\rm {\mu}m}$,  $8.6\,{\rm {\mu}m}$,  $11.3\,{\rm {\mu}m}$, $\rm H_2~S(3)$ 9.66$\,\rm{\mu}m$ and [\ion{S}{IV}] 10.5$\,\rm{\mu}m$, in the wavelength range containing the silicate feature. 
On the other hand, LEDA~1712304 has one of the most featureless continua except the silicate features in the AGN samples, and thus is expected to be one of the purest AGN-dominated galaxy in the AGN samples.
In Fig.~\ref{smoothed}, for the purpose of investigating difference, if any, in the absorption profile, we also show the optical depth profiles of the silicate features for only the AGN samples with deep silicate absorption features (${\tau}_{\rm sil}>2$), which are normalized by the optical depths averaged over the wavelength range of 9.8--10.3$\,{\rm {\mu}m}$ that include no strong lines.
Fig.~\ref{smoothed} shows that the silicate absorption profiles thus normalized are notably similar from galaxy to galaxy, although their $L_{\rm IR}$ values are much different.
And yet we find that the wings of the profiles vary on the shorter wavelength side.
At around 9$\;{\rm {\mu}m}$, the wing of LEDA~1712304 is shifted significantly toward longer wavelengths than those of many other AGNs, especially those with $L_{\rm IR}<10^{12}\,L_{\odot}$, which can be explained by differences in the compositions of amorphous olivine (e.g., the difference between models 2 and 3, see the bottom panel in Fig.~\ref{spec}) and/or the crystallinity (the difference between models 3 and 7).

\begin{figure}
   \centering
   \includegraphics[width=8cm,clip]{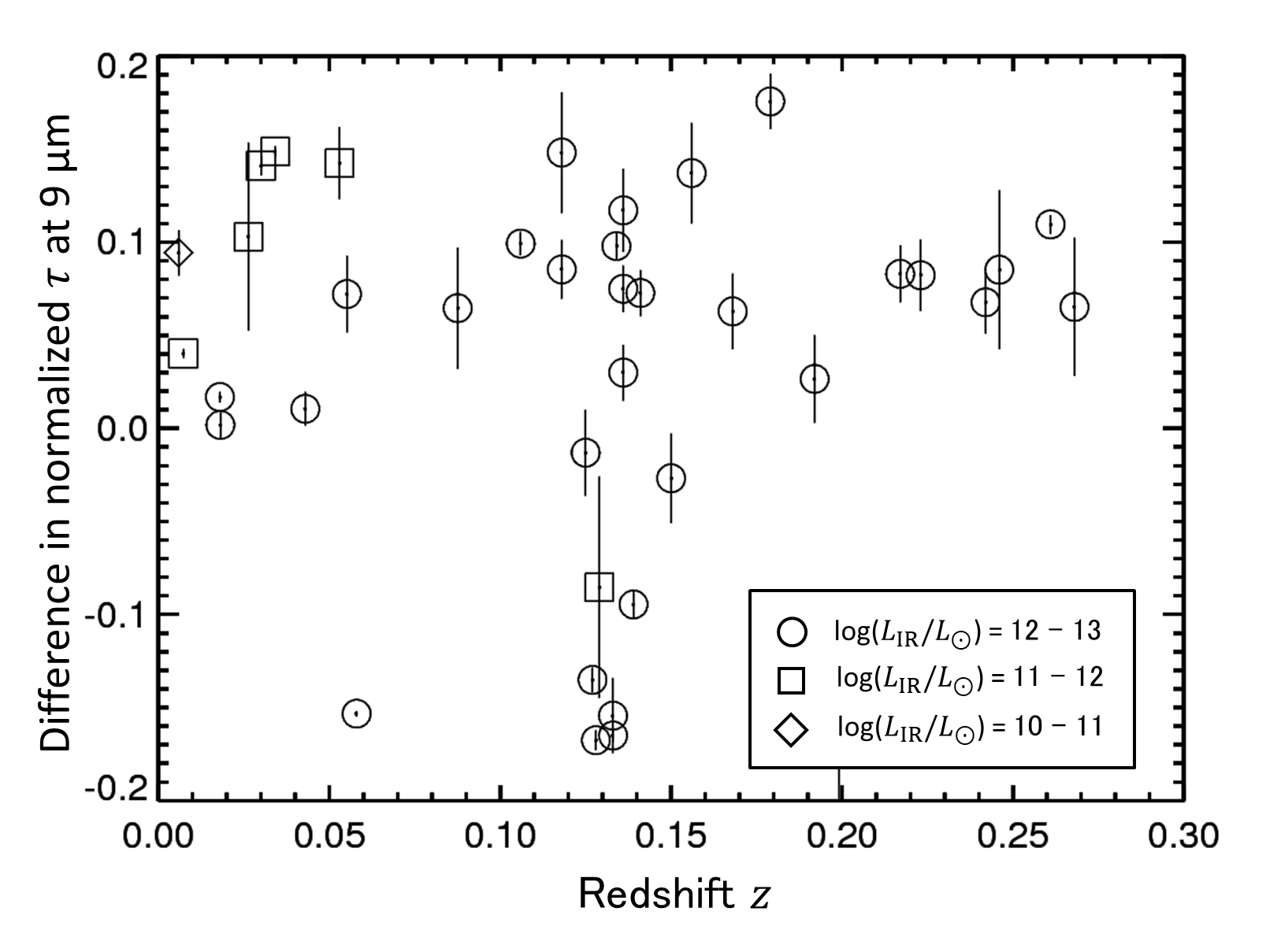}	
   \caption{Difference in the normalized optical depth at around $9\,{\rm {\mu}m}$ from that of LEDA~171304 as a function of the redshift, which is calculated by averaging the differences at 8.9--9.1$\,{\rm{\mu}m}$. A larger value indicates a deeper silicate feature at around $9\,{\rm {\mu}m}$. Circles, squares and diamonds correspond to the galaxies with $L_{\rm IR}=10^{12}$--$10^{13}\,L_{\odot}$,  $10^{11}$--$10^{12}\,L_{\odot}$ and $10^{10}$--$10^{11}\,L_{\odot}$, respectively.}	 
   \label{difference9um}%
\end{figure}

The above variations at around 9$\;{\rm {\mu}m}$ could be produced by the PAH 7.7 and $8.6\,{\rm {\mu}m}$ features filling the silicate absorption.
Since the galaxies are located at various distances, the spectroscopic apertures do not sample the same physical scales of the nuclei and thus the PAH emission from host galaxies with larger distances is expected to contaminate the spectra more strongly.
We explore this possibility in Fig.~\ref{difference9um}, where the variations at around $9\,{\rm {\mu}m}$ are plotted as a function of the redshift.
In the figure, the contamination of the PAH features from host galaxies would produce a decreasing trend with the redshift. 
However, Fig.~\ref{difference9um} does not show any clear dependence on the distance and therefore we conclude that the aperture effects with the different distances do not make an appreciable contribution to the variations of the silicate absorption profiles. %if the variations at around 9$\;{\rm {\mu}m}$ are produced by the PAH contaminations. %r=0.045}

\subsection{Silicate dust properties in LEDA~1712304}

%% galaxy evolution
%%major merger → circumnuclear starburst → concentration → quenching and PAH destruction → nuclear star old star
Such a deep silicate absorption calls for the presence of a strong dust temperature gradient \citep{Imanishi2007}, since both absorber and emitter in the mid-IR are of dust origin; the absorbing silicate dust is considered as cooler dust in the outer region of the AGN torus or interstellar dust in the host galaxy, while the strong continuum emission at around 10$\,{\rm{\mu}m}$ is attributed to hotter dust close to the AGN core. Here we consider the possibility that the absorbing dust is of interstellar origin in LEDA~1712304. As shown in Table~\ref{table:spec_para}, we estimated the column density of the absorbing dust with the spectral fitting, from which we calculate the whole mass of the silicate dust assuming the size of the galaxy, $\sim15\,{\rm kpc}$ \citep[SDSS DR6;][]{Adelman2008}\footnote{The size of LEDA~1712304 is estimated by averaging the isophotal radii of the major and minor axes in the SDSS $\it r$ band image.}, as the size of the absorbing dust layer. 
The resultant silicate dust mass is $9\times10^8\,M_{\odot}$, which is much larger than the warm and cold dust masses in Table~\ref{table:sed}. 
On the other hand, assuming that the size of the absorbing dust layer is $100\,{\rm pc}$, typical of a circumnuclear dusty torus, the silicate dust mass is estimated to be $4\times10^4\,M_{\odot}$, which is reasonable as compared with the warm dust mass in Table~\ref{table:sed}.
Therefore the absorbing silicate dust is likely of not interstellar but circumnuclear origin, which is present in the outer region of the AGN torus. % + PAHが見られず、星形成が起こっていないことから、中心のAGNにものが落ち込んで → iceの後ろ
%This is consistent with the observational studies that mid-IR sources in ULIRGs are concentrated in the nuclei \citep[e.g., ][]{Soifer2000}.

%\citet{} suggested a galaxy evolutionary scenario that merger driven a starburst, after that AGN is activated, and then the galaxy evolve to QSO or Elliptical. Thus LEDA1712304 may be in the phase that .%後期段階be estimated

%The IR spectrum also shows the featureless continuum at 2.5-4.5$\,{\rm{\mu}m}$ and 5.5-8$\,{\rm{\mu}m}$. The interstellar PAH and ice features are difficult to be detected, since the hot dust emission from the AGN torus dominates. The non-detection of PAH features implies a destruction in the center by X-ray plasma from the AGN or a degeneration high temperature. %cold dustに対するLPAHが典型的な値に対して、低ければ、壊れているなどと議論可能。
%ice features AGN intensity cold dustの量の兼ね合い
%The star formation rate (SFR) can be estimated by $L_{\rm IR}$ in Table~\ref{table:sed} \citep{}, and the resultant SFR is　7.6 $M_\odot$/yr. %もともとないとかは？
%The ice features were not also detected, outer torus higher than sublimation temperature. minor merger dust poor consistent. other dust-obscured galaxies which have deep silicate absorption are variable strength. This is the power of the AGN to total dust  

%buried AGN dust NLRない　polar dust
%So we can consider the dust concentration driven by a minor merger

%% processing of dust 
The silicate feature of the AGN torus dust in LEDA~1712304 is different in the profile from that of the astronomical silicate. 
The result of the IR spectral fitting indicates that the difference in the silicate features is attributed to either composition, the physical structure of dust, or both. 
The best-fit model (model 7) needs amorphous olivine as a dominant composition, which is consistent with the result of \citet{Spoon2006} for heavily obscured ULIRG nuclei.
Based on the fact that the main composition of dust in the circumstellar dust shells around late M stars is also amorphous olivine \citep{Roche2007}, we speculate that the grains of amorphous olivine generated in the old stars may have fallen into the galactic center as induced by a galaxy interaction, for example, which can explain amorphous-olivine rich dusty torus around the AGN.
%%どのあたりから落ちてきてtimescale的にはどうか？

\citet{Spoon2004} suggested the presence of an unobscured emission component to explain the flat spectrum at the bottom of the silicate feature of a heavily obscured AGN.
Indeed, the model which considers an absorption-free continuum (model 4) gives a better fit to the spectrum of LEDA~1712304 than that which only considers the obscured circumnuclear hot dust (model 2).
On the other hand, the even better-fit result of the IR spectral fitting is the porous amorphous olivine (model 5 and 7).
The porosity of the silicate dust may be caused by coagulations or processing in the circumnuclear environments.  
%\cite{Li2008} and \citet{Smith2010} reported that silicate emission features in type-1 AGNs are also fitted well with porous amorphous olivine. 
The result of the IR spectral fitting also suggests that the dusty torus around the AGN in LEDA~1712304 is likely to contain the crystalline silicate.
\citet{Kemper2004} reported that the crystallinity of the diffuse interstellar silicate dust in our Galaxy is smaller than $2.2\%$ and the absence of crystalline silicate is explained by an amorphization process caused by cosmic-ray particle bombardment which occurs on a timescale significantly shorter than the destruction timescale.
The crystallinity of silicate dust in the AGN torus in LEDA~1712304 is ${\sim}4\%$ from the IR spectral fitting, which is larger than that of the diffuse interstellar dust in our Galaxy.
Therefore the silicate dust in the AGN torus in LEDA~1712304 is likely to be relatively fresh, possibly formed in the circumnuclear dense environments.

\section{Conclusions}
We detect a deep ($\tau_{\rm sil} \sim 2.3$) absorption feature due to silicate grains at around 10$\,{\rm{\mu}m}$ in the AKARI/IRC near- to mid-IR spectrum of the nearby heavily obscured galaxy LEDA~1712304.
The spectrum also shows a steep near- to mid-IR slope, indicating that LEDA~1712304 possesses a heavily obscured AGN. 
The IR luminosity and stellar mass of LEDA~1712304, $L_{\rm IR}=(5\pm1){\times}10^{10}\,L_{\odot}$ and $M_{\rm star}=(2.7\pm0.8)\times10^{9}\,M_{\odot}$, are notably low compared with other heavily obscured AGNs which show deep silicate absorptions. %ここまで結果
Thus LEDA~1712304 may be a rare galaxy showing low $L_{\rm IR}$, low $M_{\rm star}$ and yet large ${\tau}_{\rm sil}$.
On the other hand, we find that the spectral profile of the silicate feature in LEDA~1712304 is similar to those of the other AGN samples as a whole, but significantly different in the wing on the shorter wavelength side, which can be explained by difference in the compositions and/or the crystallinity.
 %ellipticalのとこ
%楕円銀河へのevolutionary scenarioとして知られるmergerを考えると、軽い銀河ほどmergerが起こりにくいと考えられ、rare性を説明できる。

%absorbing silicate dust はnot interstellar but circumnuclear originであることはsurely。FIRのemissionから見積もられるdust mass とτから見積もられるdust massの間のconsistency を考えると、
The absorbing silicate dust in LEDA~1712304 is not of interstellar origin but of circumnuclear origin, since the dust mass estimated from $\tau_{\rm sil}$ and the size of host galaxy is much larger than that estimated from the far-IR emission.
From the IR spectral fitting, the main composition of the circumnuclear silicate dust in LEDA~1712304 is amorphous olivine, which is consistent with the previous studies \citep[e.g.,][]{Spoon2006}. %古い星起源ではないか？
In addition, the best-fit model of the IR spectral fitting calls for the porosity and the crystallinity of the silicate dust, which imply the dust coagulation or processing and recent dust formation, respectively, in the circumnuclear environments.

\begin{acknowledgements}
We thank the referee for carefully reading our manuscript and giving us helpful comments. This work is based on observations with AKARI, a JAXA project with the participation of ESA, with the Spitzer Space Telescope, which is operated by the Jet Propulsion Laboratory, California Institute of Technology under a contract with NASA, using the NASA/IPAC Infrared Science Archive and with {\it Herschel}, which is an ESA space observatory with science instruments provided by European-led Principal Investigator consortia and with important participation from NASA.
The IRSF project was financially supported by the Sumitomo foundation and Grants-in-Aid for Scientific Research on Priority Areas (A) (Nos. 10147207 and 10147214) from the Ministry of Education, Culture, Sports, Science and Technology (MEXT). 
The operation of IRSF is supported by Joint Development Research of National Astronomical Observatory of Japan, and Optical Near-Infrared Astronomy Inter-University Cooperation Program, funded by the MEXT of Japan.

%      Part of this work was supported by the German
%      \emph{Deut\-sche For\-schungs\-ge\-mein\-schaft, DFG\/} project
%      number Ts~17/2--1.
\end{acknowledgements}

   \bibliographystyle{aa} % style aa.bst
   
 % your references Yourfile.bib

\end{document}